\documentclass[%
 reprint, pdftex
amsmath,
amssymb,
aps,
pra
showpacs
]{revtex4-1}

\usepackage{graphicx}
\usepackage{dcolumn}
\usepackage{bm}
\usepackage[usenames,dvipsnames,svgnames,table]{xcolor}
\usepackage[normalem]{ulem}
\usepackage{upgreek}

\begin{document}

\preprint{APS/123-QED}

\title{Dynamics of bad-cavity enhanced interaction with cold Sr atoms for laser stabilization}%

\author{S. A. Sch\"{a}ffer}
\email{schaffer@nbi.dk}
\affiliation{Niels Bohr Institute, University of Copenhagen, Blegdamsvej 17, 2100 Copenhagen, Denmark}
\author{B. T. R. Christensen}%
\affiliation{Niels Bohr Institute, University of Copenhagen, Blegdamsvej 17, 2100 Copenhagen, Denmark}
\author{M. R. Henriksen}
\altaffiliation[Also at ]{SPOC, Technical University of Denmark,  DTU Fotonik, Ørsteds Plads, building 343, 2800 Kgs. Lyngby, Denmark}%
\affiliation{Niels Bohr Institute, University of Copenhagen, Blegdamsvej 17, 2100 Copenhagen, Denmark}
\author{J. W. Thomsen}%
\affiliation{Niels Bohr Institute, University of Copenhagen, Blegdamsvej 17, 2100 Copenhagen, Denmark}

\begin{abstract}
Hybrid systems of cold atoms and optical cavities are promising systems for increasing the stability of laser oscillators used in quantum metrology and atomic clocks. In this paper we map out the atom-cavity dynamics in such a system and demonstrate limitations as well as robustness of the approach. We investigate the phase response of an ensemble of cold strontium-88 atoms inside an optical cavity for use as an error signal in laser frequency stabilization. With this system we realize a regime where the high atomic phase-shift limits the dynamical locking range. The limitation is caused by the cavity transfer function relating input field to output field. However, the cavity dynamics is shown to have only little influence on the prospects for laser stabilization making the system robust towards cavity fluctuations and ideal for the improvement of future narrow linewidth lasers.
\end{abstract}

\maketitle

\section{Introduction}
Optical atomic clocks have undergone an immense development, and are continuously improving, with increased stability and accuracy every year \cite{Bloom, Hinkley, LeTargat, Ushijima}. The ability to reach exceedingly high accuracies within a reasonable time is made possible by the correspondingly huge effort to bring down the frequency noise in ultra stable laser sources \cite{Kessler,Bishof,Hafner,Matei}.

The full potential of the high Q factor atomic transitions used in many optical atomic clocks can be reached only through improvements in the stability of the interrogation laser.
Traditionally such interrogation lasers are stabilized to highly isolated optical reference cavities. This stabilization method is mainly limited by thermal fluctuations in the optical coating, mirror substrate and cavity spacer \cite{Matei,Numata,Cole,Kessler_2} demanding considerable experimental effort in order to construct cryogenically cooled mono-crystalline cavities and crystalline mirror coatings \cite{Kessler, Matei}. Several new approaches are being pursued in the so-called bad cavity regime \cite{Kuppens}, in order to significantly suppress thermally induced length fluctuations. They use a combination of narrow linewidth $\updelta\nu$ atoms and optical cavities. These atomic systems have transitions at optical frequencies $\nu$, with strongly forbidden transitions resulting in high Q factors, $Q=\frac{\nu}{\updelta\nu}$. By exploiting the high Q factor of the atomic transitions and using cavities with comparatively low Q factors the systems are far less sensitive to thermal fluctuations of the cavity components, and the experimental requirements are simplified. In these approaches active as well as passive atomic systems have been suggested \cite{Chen, Meiser, Maier, Kazakov, Martin, Schaffer}. The active atomic systems are optical equivalents of the maser, relying on co-operative quantum phenomena such as superradiance or superfluorescence of atoms inside the cavity mode. Several pioneering experiments have already demonstrated lasing under such conditions \cite{Bohnet,Zhuang,Pan,Norcia,Norcia2}. In the passive approach the atom-cavity system is used as a reference for laser stabilization where the narrow linewidth atomic transitions are interrogated inside an optical cavity. One proof-of-principle approach to this is based on using the NICE-OHMS technique \cite{NICE-OHMS,NICE-OHMS2} for generating sub-Doppler dispersion signals \cite{Axner}. This has shown promising results for laser stabilization that could be able to compete with traditional cavity stabilization techniques \cite{Westergaard,Christensen,Tieri}. 

By employing an optical cavity the coupling between atoms and optical field is improved by a factor of the cavity finesse, which significantly increases the total phase-shift experienced by the optical field. As the total phase-shift is increased, however, this limits the frequency range of linear behavior and thus the dynamical range of a servo locking the laser frequency. Additionally, the cavity servo response time might limit the signal quality if the condition of constant laser-to-cavity resonance must be strictly met.

In this paper we show experimentally that the large total phase shift of the system not only improves the resonance slope, but also distorts the dispersion signal off atomic resonance. This becomes relevant for the interest of servo optimization in such a system \cite{Leroux} as it can limit the dynamical range of a servo lock. We show that this distortion originates from the transfer function of the cavity itself, and thus cannot be circumvented. We have realized a system with a theoretically attainable shot noise limited laser linewidth of $\Delta\nu\approx40$~mHz, possibly allowing laser performance at the level of the state-of-the-art reported values \cite{Kessler,Bishof,Hafner,Matei}. We use the system to map out the dynamical range and investigate the consequences of an imperfect cavity servo, which causes a mismatch of the cavity resonance with respect to the laser frequency. Due to the bad cavity regime much looser bounds on the cavity resonance are allowed, as expected. This opens the possibility of using cavities with quasi-stationary lengths, and simultaneously underlines the insensitivity to cavity fluctuations.

\section{Experimental system}
\begin{figure}[h]
	\includegraphics[width=\columnwidth]{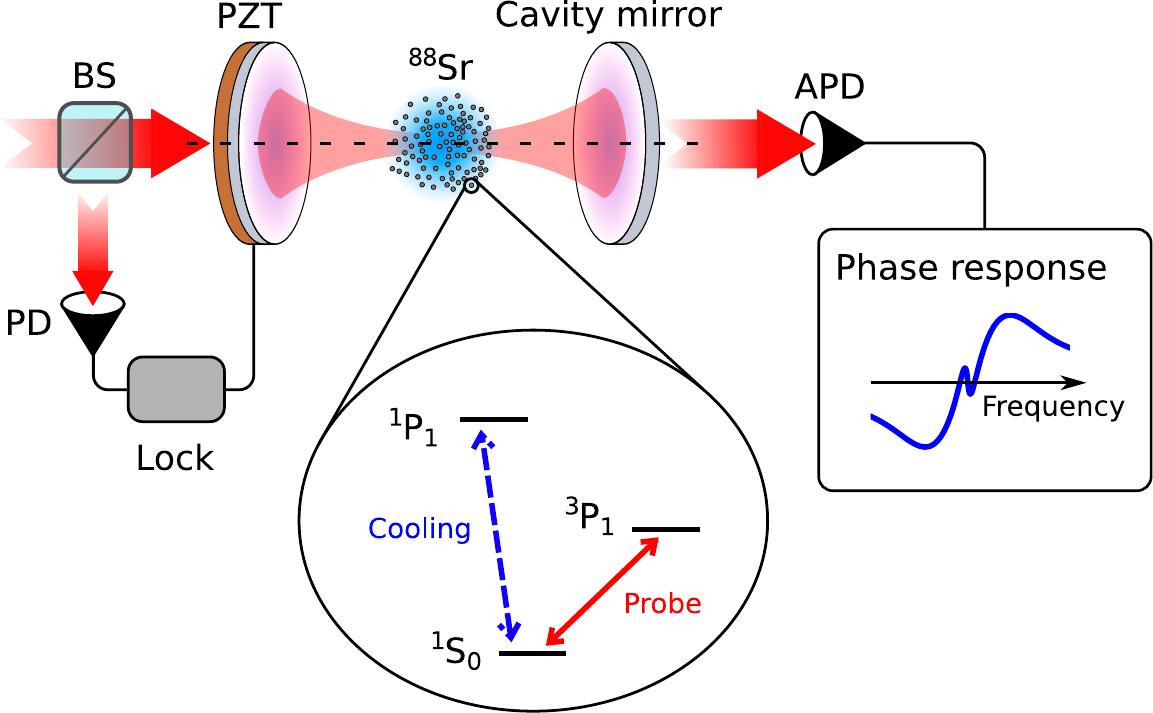}
	\caption{Experimental system. The probe laser has a single carrier frequency as well as sidebands detuned an integer number of the cavity free spectral range (FSR). The light is coupled into the cavity by adjusting the cavity length to ensure resonance between the atom-cavity system and the probing laser at all times. This resonance is ensured by a Pound-Drever-Hall lock from the reflected light using the beamsplitter (BS), photodetector (PD) and locking circuit acting on the piezo-electric element (PZT) on one of the cavity mirrors. The sidebands do not interact with the atoms inside the cavity and act as references in the subsequent heterodyne measurement of the transmission signal on the fast avalanche photodetector (APD). The cooling transition for the MOT and the probe transitions are shown. During measurements the cooling light (blue) is turned off. \label{setup}}
\end{figure}

The experimental system investigated here consists of an ensemble of cold strontium-88 atoms cooled to a temperature of $T \approx 5$~mK. The atoms are trapped in a Magneto-Optical Trap (MOT) at the center of a TEM$_{00}$ Gaussian mode of an optical cavity, see FIG.~\ref{setup}. The cavity has a finesse of $F=1240$ and a linewidth of $\kappa=2\pi\cdot 630$~kHz at $\lambda=689$~nm. A laser beam probing the narrow $(5s^2)$ $^1$S$_0 \rightarrow\,(5s5p)\, ^3$P$_1$ transition of $^{88}$Sr is coupled into the cavity mode, and the cavity resonance is locked to the probe laser frequency at all times.

Before entering the cavity the probe light is phase-modulated using a fiber-coupled Electro-Optical Modulator (EOM) in order to perform heterodyne detection of the transmitted signal. The modulation frequency is equal to the free spectral range (FSR) of the cavity resulting in sidebands at $\omega_0 \pm j\Omega$ for integer $j$ and  $\Omega=2\pi\cdot 781.14$~MHz. The sidebands are far detuned with respect to the $(5s^2)$ $^1$S$_0 \rightarrow\,(5s5p)\, ^3$P$_1$ transition having a linewidth of $\gamma_{\text{nat}}=2\pi\cdot7.5$~kHz, and the interaction between the sidebands and the atoms can thus be neglected. This system is interrogated using a heterodyne measurement (the NICE-OHMS technique \cite{NICE-OHMS,NICE-OHMS2}) in order to extract the dispersion signal of the atom-cavity system, which can be used as an error signal to lock the probe laser frequency to resonance with the atoms. We operate in the bad cavity regime where any cavity fluctuations are suppressed in the atom-cavity signal by a factor of $\frac{\kappa}{\gamma_{\text{nat}}}$, here about $100$.

The field transmitted through the cavity is split and simultaneously recorded on a low bandwidth ($50$~MHz) photodiode and a high bandwidth ($1$~GHz) avalanche photodetector (APD). The low bandwidth signal records the total transmission intensity of the cavity. The high bandwidth signal is filtered around the modulation frequency $\Omega$ and demodulated in order to record the atom induced phase shift of the sideband relative to the carrier frequencies.

The measurements are performed in a cyclic operation as the intense cooling light of the MOT results in an AC Stark shift of the $^3$P$_1$ level and washes out coherence of the probing transition. The cooling light is thus shut off before each measurement, and the probing light then recorded for an interrogation period of $100~\upmu$s. At this timescale the probing laser has a linewidth of $\Gamma_{\text{l}}=2\pi\cdot800$~Hz which is much narrower than the natural linewidth of the probing transition $\gamma_{\text{nat}}=2\pi\cdot7.5$~kHz. This transition linewidth places us deep in the bad cavity regime, where the cavity linewidth is much broader than the atomic linewidth $\kappa\gg\gamma$. This means that the system is much less sensitive to variations in the cavity resonance frequency which can originate from, e.g., temperature fluctuations in the cavity components

Only a single measurement is performed before reloading the trap with new atoms, since atom loss due to the finite temperature of the atoms becomes measurable after $500~\upmu$s. This results in a cyclic operation where the dispersion is measured only for a single frequency detuning of the interrogation laser at a time. Varying the loading time of the MOT allows control over the atom number and typically ranges from $50$~ms to $800$~ms for intra-cavity numbers of $N=2\cdot10^6$ to $N=4\cdot10^7$.

\section{Theory of measurement}
We investigate theoretically a system consisting of an ensemble of $N$ atoms coupled to a single mode of an optical cavity in order to describe the experimental system presented in this work. The NICE-OHMS technique as it is used here relies on the transmitted signal of the atom-cavity system, and is a heterodyne measurement between the carrier laser frequency and its sidebands. The input laser field before the cavity can then be described by
\begin{equation}
E_{\text{in}}=E_0\sum_{j=-\infty}^{\infty}J_{j}(y)e^{i(\omega_l+j\Omega)t},
\end{equation}
where $E_0$ is the amplitude of the electric field, $J_j(y)$ is the $j$'th order Bessel function of the first kind, with the modulation index $y$. The laser carrier frequency is $\omega_l$, whereas $\Omega$ is the modulation frequency applied in the EOM.

The interaction of the light with the atom-cavity system may be described by using a Born-Markov master equation as described in appendix \ref{Appendix:theory} following \cite{Westergaard, Tieri}. The approach is based on a many-particle Hamiltonian $\hat{H}$ and a derived set of complex Langevin equations that includes the Doppler effect from the finite velocity of the atoms.

Classically we may relate the input and output fields with a complex transfer function $\chi\left[\theta(E_{\text{in}})\right]$. The field-dependent complex atomic phase experienced by the light when interacting with the atom-cavity system $\theta(E_{\text{in}})$ is found by means of the full quantum mechanical theory of appendix \ref{Appendix:theory}. In order to cast the behavior of our system in terms of measurable quantities, we assume that the relation between the quantum mechanical phase $\theta(E_{\text{in}})$ and the measured output power can be described by a linear model such that $E_{\text{out}}=\chi(\theta) E_{\text{in}}$. We then insert the theoretical value of $\theta(E_{\text{in}})$ into the transfer function.

Here we are mainly interested in the properties of such a transfer function. By increasing the finesse of the cavity with respect to the numbers reported in \cite{Christensen} we enable the system to move between a low-phase-shift regime and a high-phase-shift regime.

\begin{figure}[h!]
	\includegraphics[width=0.98\columnwidth]{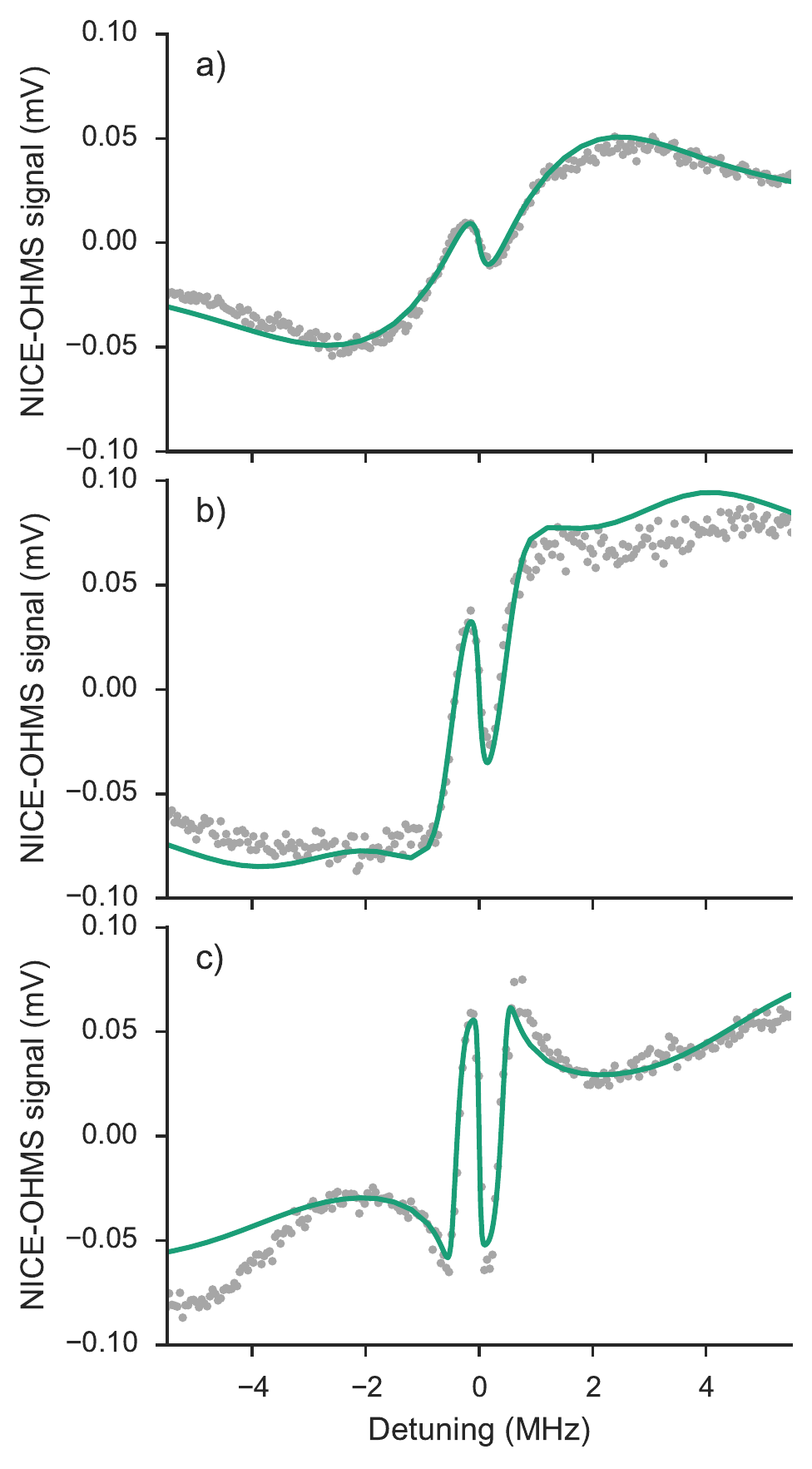}
	\caption{Phase-response of the atom-cavity system using the NICE-OHMS technique. We show the transition from low to high phase-response by changing the number of atoms interacting with the cavity mode. Here the gray dots indicate recorded data, whereas the full green curve is a theoretical calculation using the experimental parameters. As the absolute phase increases the transfer function reaches its maximal value and first flattens, then inverts the signal slope. {\bf a)} Cavity atom number of $N=3.8\cdot10^6$, and a temperature of $T=16$~mK. Due to the short loading time used the atoms are slightly warmer here. 
	{\bf b)} Cavity atom number of $N=1.4\cdot10^7$, and a temperature of $T=13$~mK. Distortion in the dispersion signal is evident from $\pm 1$~MHz to $\pm 4$~MHz. {\bf c)} Atom number of $N=4\cdot10^7$, and a temperature of $T=13$~mK. Here we clearly see a slope inversion initiating at $\pm 1$~MHz. The maximal absolute value of the dispersion is not constant over the whole scan, as it depends on the absorption which is itself dependent on the detuning. This also causes the dispersion to retain a large absolute value for relatively high detuning.  \label{transition}}
\end{figure}

FIG.~\ref{transition} shows typical dispersion-scans where the theoretical model incorporating a cavity transfer function, see \cite{Christensen}, has been plotted using the known experimental parameters. The dispersion signal serves as an error signal for all values of the atom number $N$, but is distorted when detuned from resonance at higher values of $N$. This distortion is not caused by the atomic phase-response itself, but rather by the classical conditions of the transfer function imposed by the cavity.

\subsection{Dispersion signal}

Only a single frequency component of the modulated light, namely the carrier component $j=0$,  interacts with the atoms. This means that we can simplify the description of our system by defining a transfer function for each frequency component $j$ of the light as it passes through the cavity \cite{Riehle}
\begin{equation}\label{eq:transfer}
\chi_j = \frac{Te^{i\phi_j}}{1-Re^{2i\phi_j}},
\end{equation}
where $T$ ($R$) is the power transmission (reflectivity) of a single cavity mirror, and $\phi_j$ is the complex phase experienced by the $j$'th component of the interrogation laser. We assume identical mirrors with no losses. The real part of the transfer function corresponds to the transmitted amplitude of the $E$-field in the system, while the imaginary part corresponds to the dispersion. Due to energy conservation the absolute-squared value of the complex transfer function cannot exceed one, $|\chi|^2\leq1 $, for a system with no gain or frequency conversions. This classical condition thus imposes a maximal value on the dispersion signal which is independent on the nature of the phase-delay inside the cavity. We can describe the complex phase for any sideband component as simply the phase-shift experienced by a single-passage interaction with the cavity $\phi_{j}=\phi_{\text{cav}}^j$ for $j\neq0$, while the carrier component of the light experiences the atomic phase as well
\begin{equation}
\phi_0=\phi_{\text{cav}}^0 + \phi_D + i\phi_A,
\end{equation}
where $\phi_D$ and $\phi_A$ are the phase components caused by atomic dispersion and absorption from a single passage of the cavity. In the case of a medium with no gain, we have $\phi_A \geq0$. The cavity phase-shift is given by $\phi_{\text{cav}}^j=\phi_{\text{cav}} + j\pi$, and the cavity locking conditions of the experiment defines $\phi_{\text{cav}}$. 

The output field can now be expressed by a superposition of frequency components and corresponding transfer functions
\begin{equation}
E_{\text{out}}=E_0 \sum_{j=-\infty}^{\infty} J_{j}(y)\chi_j e^{i(\omega_l+j\Omega)t},
\end{equation}
where $E_0$ contains any overall phase. By recording the intensity on a photodetector we can filter out the beat signal between sideband and carrier by demodulating at the modulation frequency $\Omega$. By optimizing the phase of the demodulation signal to record the imaginary part of the transfer function and subsequently pass the signal through a $2$~MHz low-pass filter we obtain a DC signal
\begin{equation}\label{eq:S0_1}
S_\Omega\propto 2i|E_0|^2 J_0(y)J_1(y)(\chi_0\chi_1^*-\chi_0^*\chi_1),
\end{equation}
which is a purely real number. We have only included up to second order sidebands, and used $\chi_{j}=(-1)^{|j|-1}\chi_1$ for $j\neq0$. Higher order sidebands are negligible for modulation indices up to $y\simeq1$.

If we assume that the system is in a steady state the cavity locking condition dictates that the cavity is on resonance with the carrier frequency at all times, corresponding to that used in \cite{Christensen}. This gives us 
\begin{equation}
\phi_{\text{cav}} + \phi_D = n\pi \label{eq:phicav}
\end{equation}
for integer $n$. The complex transfer function of the carrier then becomes solely dependent on the absorption

\begin{equation}
\chi_0 = \frac{Te^{-\phi_A}}{1-Re^{-2\phi_A}},
\end{equation}
whereas the sideband transfer functions have the phase information of the atomic interaction written onto them by the cavity lock
\begin{eqnarray}
\phi_{j}&=&\phi_{\text{cav}}^j =\phi_{\text{cav}} + j\pi \quad \text{for}\ j\neq0\nonumber\\
&=& n\pi - \phi_D + j\pi.
\end{eqnarray}
Ignoring an overall sign from $e^{i n\pi}$ we get
\begin{equation}
\chi_{j} = \frac{Te^{i(j\pi-\phi_D)}}{1-Re^{2i(j\pi-\phi_D)}} \quad \text{for}\ j\neq0.
\end{equation}
Since $\chi_0$ is purely real we can write the signal as
\begin{equation}
S_\Omega\propto J_0(y)J_1(y)\chi_0\textrm{Im}(\chi_1). \label{eq:S0}
\end{equation}
We are thus particularly interested in the properties of the imaginary part of the transfer function if we wish to understand the behavior of our error signal.

\subsection{Transfer function properties}
Having understood the behavior of our system we can now investigate why we see the folding behavior depicted in FIG.~\ref{transition} \textbf{b)} and \textbf{c)}. If we ignore the origin of the phases it is clear that a cavity transfer function such as the one in Eq.~\ref{eq:transfer} must have a periodicity of $2\pi$ as a function of the phase shift experienced by the light inside the cavity. In connection with locking of the laser to an atom-cavity system we are mainly interested in the phase slope around atomic resonance where the absolute phase is zero, but the phase slope can be very large. 

\begin{figure}[h]
	\includegraphics[width=\columnwidth]{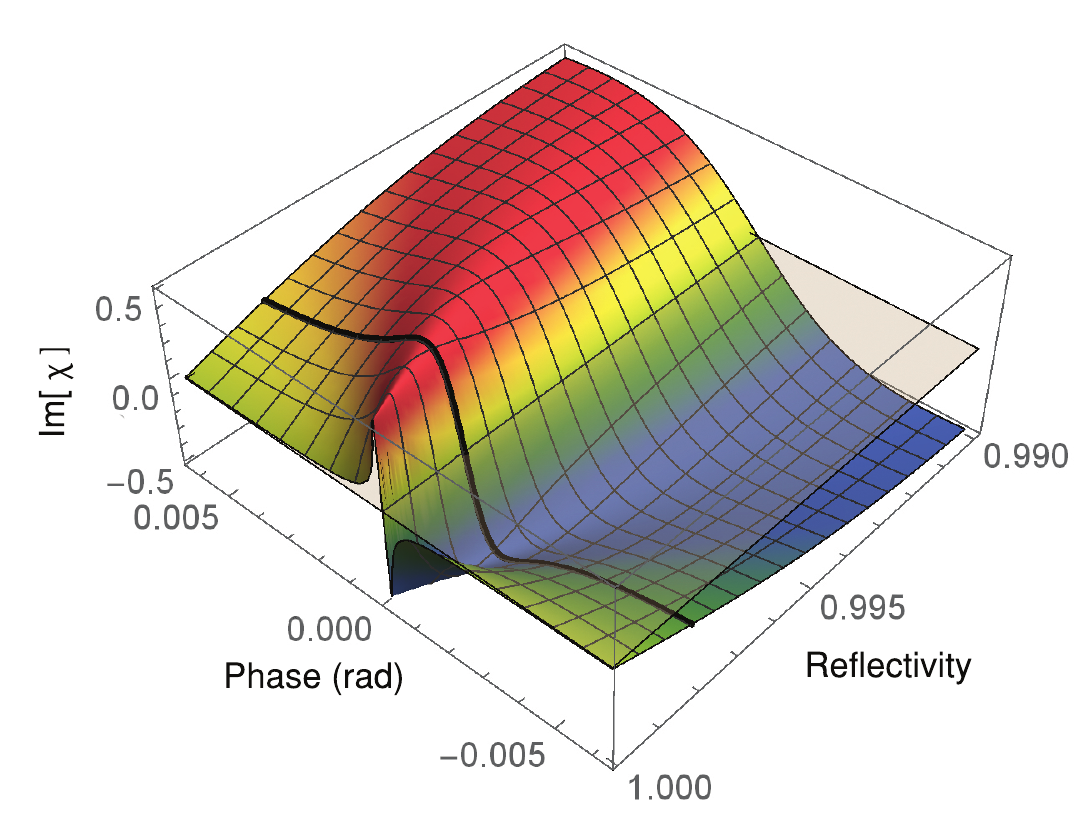}
	\caption{The imaginary part of a transfer function as given in Eq.~\ref{eq:transfer} as a function of phase $\phi$ and single mirror power reflectivity $R$. Here we have assumed that the phase $\phi$ is purely real, and that the cavity is symmetrical. Close to zero phase the function is approximately linear, and the values of the phase and the transfer function are proportional. As the phase increases, however, a maximal value for the transfer function is reached, and the transfer function slope is inverted. The black line indicates the mirror reflectance of our system ($R=0.998$) and the transparent plane indicates $\chi=0$. \label{transferfunc}}
\end{figure}

Very close to atomic resonance, the transfer function is proportional to $\sin(\phi)\approx\phi$ \cite{Christensen} and we can treat the transfer function as linear in phase. For a slightly larger frequency detuning, however, the existence of a maximal value for the transfer function results in some interesting behavior for a system with large total phase shift. In FIG.~\ref{transferfunc} the imaginary part of a phase-dependent transfer function $\chi_j$ is shown with varying single-passage phase-shift and mirror reflectivity $R$. We see that the imaginary part of the transfer function itself behaves dispersion-like for a linearly varying phase. In this figure we have assumed that there are no losses in the cavity mirrors ($T+R=1$) and that there is no absorption in the cavity $\textrm{Im}[\phi] = 0$ which would not be the case close to an atomic resonance. If the effects of absorption in the medium is taken into account, this reduces the maximal value of transfer function $|\chi|_{\text{max}}$ further. For $\textrm{Im}[\phi] = \phi_A > 0$ we will thus have $|\chi|_{\text{max}}< 1$ asymptotically decreasing towards zero as a function of $\phi_A$. As an aside, including absorption also decreases the phase slope at resonance. This slope will nevertheless still increase linearly with atom number when the saturation condition is fulfilled.

As the reflectivity of the mirrors ($R$) is increased the light is stored in the cavity for longer and thus experiences a larger total phase shift. This increases the phase slope on resonance proportionally to the finesse $F$ of the cavity and in turn leads to a decrease in the phase range where the transfer function $\chi$ is linear, see FIG.~\ref{transferfunc}. This insight tells us that the dispersion signal observed from the atom-cavity transfer function will be distorted and even change the sign of the slope for detunings at which the values of the total phase shift is large.

This sets a limit to the maximal \emph{dynamical range} that we can expect of a locking mechanism based on this dispersion signal $S_\Omega$. It results in an inversion of the dispersion slope for large absolute phase-shifts. Here the boundaries on the transfer function act to fold down the signal in a non-linear manner. While the sign of the slope is thus inverted the sign of the signal itself never changes with respect to that of the phase. The linear-phase regime decreases in size linearly towards zero as a function of the mirror reflectivity $R$ in the regime where the cavity linewidth $\kappa \ll \text{FSR}$ ($F \gg 1$). A maximal dynamical range of $\phi=\pi$ is reached for $R\lesssim0.17$. For systems with much larger atom number (and thus larger phase-shift) it could thus be an advantage to go towards lower mirror reflectivity, and thus deeper into the bad-cavity regime. This would further reduce the sensitivity to cavity perturbations. For systems using much broader atomic transitions where the cavity might naturally have lower finesse \cite{Chen}, these effects would only be visible for very large samples.

The absolute phase value at which such mirroring occurs typically increases with larger detuning from the resonance. This effect is caused by the decrease in atomic absorption for increased detuning. This causes the phase value necessary for the slope inversion of the transfer function to increase. Away from resonance the dispersion is thus highly distorted, with respect to the atomic phase, due to the functional form of the cavity transfer function.

\section{Results and Discussion}
Here we report on the phase response of the system when operating in a regime of high phase shift due to a combination of large atom numbers $N$ and high reflectance of the cavity mirrors. At small frequency detuning we see a linear scaling of the dispersion slope with respect to the phase slope, which gives us a limit on the ultimate frequency linewidth of a laser locked to such a system \cite{Martin,Christensen}. The dynamical range of a laser frequency lock to the atom-cavity system becomes limited at high absolute phase-shifts. This is caused by the characteristics of the transfer function whose behavior will then dominate over the power broadened transition linewidth $\Gamma_\textrm{\tiny p}$. We quantify this limitation and its implications for laser frequency locking. We have also investigated the effects of having a cavity resonance lock with non-optimal conditions. The modification of such locking conditions is of interest to any experimental realization of the frequency lock.

\subsection{Phase-slope and projected shot-noise limited linewidth}
In the context of locking the frequency of a laser to the atom-cavity system, we are interested in obtaining an error signal that we can use as a feed-back signal, which must have a large slope and a large signal-to-noise ratio (SNR). The first condition is limited by the physical system, and is given by the phase-slope present at resonance. The second condition is limited by the noise present in the experimental system, and is to a high degree limited by technical circumstances that may be significantly reduced. These technical contributions to the noise include residual amplitude modulation (RAM) of the laser sideband components, atom number fluctuations and noise in the detectors. Because of this fundamental difference in the two conditions, we wish to focus on the limitations set by the physical system initially - namely the phase-slope at resonance.

\begin{figure}[h]
	\includegraphics[width=\columnwidth]{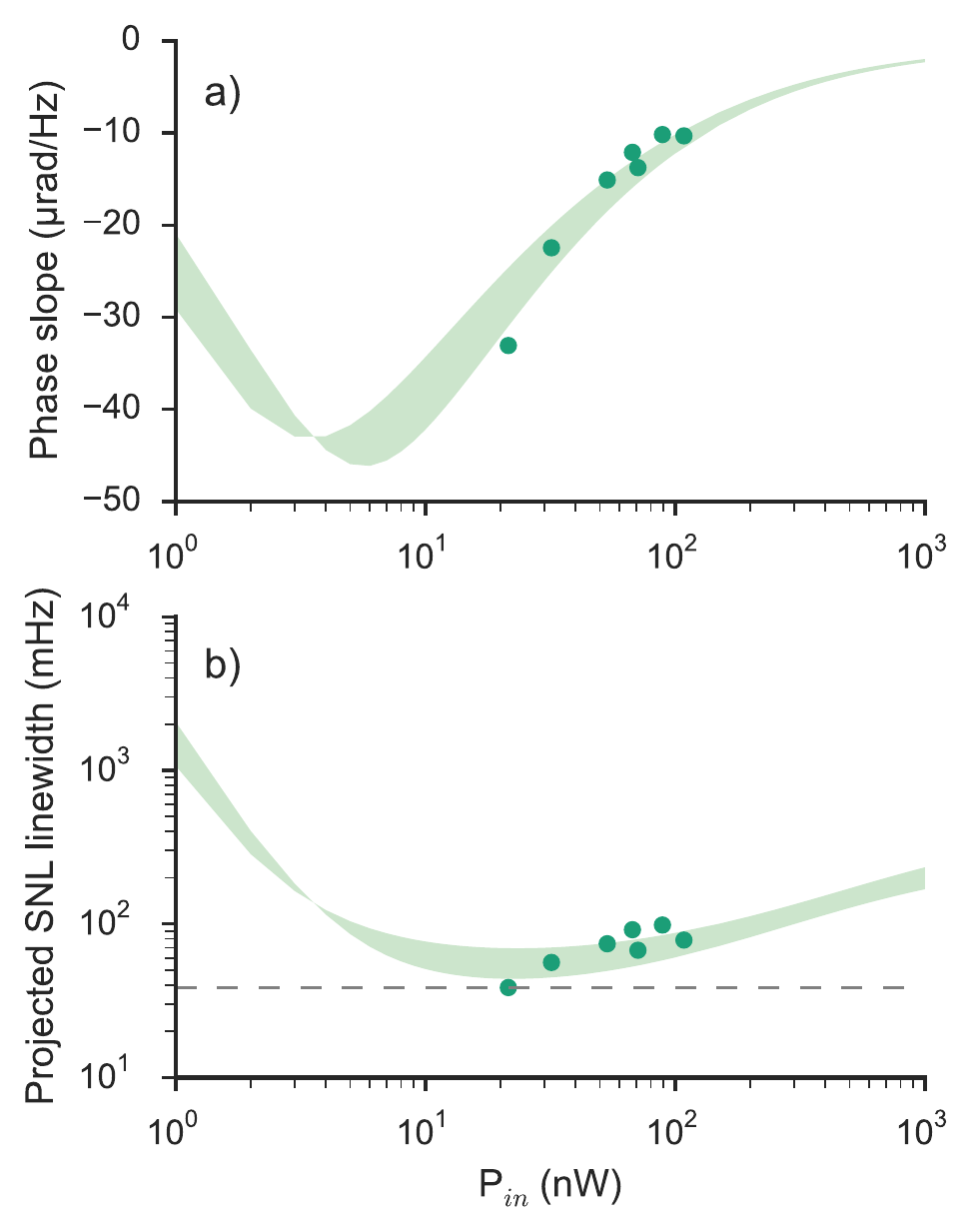}
	\caption{ {\bf a)} Semi-logarithmic plot of the slope of the atomically induced phase-slope at resonance, $\frac{d\phi}{d\nu}$, as a function of input power. The shaded area is the range between theoretical predictions for experimental parameters at a temperature $T=3.6 \pm 1$~mK and $N=2.7\cdot10^7 \pm 5\cdot10^5$ atoms overlapping with the cavity mode. This represents the uncertainty in atom number due to shot-to-shot variations, as well as the uncertainty in temperature mainly caused by power fluctuations of the cooling laser. For a number of different input powers we measured the dispersion and found the phase-slope of the theoretical fit. The uncertainty in input power $P_{\text{in}}$ is less than or equal to the dot-size. {\bf b)} Projected Shot-Noise Limited (SNL) linewidth achievable for the system under ideal circumstances. The dots are calculated values corresponding to the slopes found in \textbf{a)}. The single-sideband-to-carrier power ratio is $0.5$. We see a minimal SNL linewidth of $\Delta\nu\approx40$~mHz which is comparable to the current state-of-the-art results.\label{slope}}
\end{figure}

In FIG.~\ref{slope}~\textbf{a)} the slope of the atomic induced phase shift at resonance is plotted as a function of the input power on a logarithmic scale for $N=2.7\cdot10^7$. It was shown in \cite{Westergaard} that the slope at resonance scales linearly with the number of atoms $N$ in the cavity mode. This is still the case in our regime of $N\approx1-5\cdot10^7$ and $P_{\text{in}}\simeq100$~nW \cite{ChristensenPHD}, and we will thus focus on the strongly non-linear scaling with laser power here. This scaling was shown for a cavity finesse of $\mathcal{F}=75$ in \cite{Christensen}. Here we show results for a system with finesse of $\mathcal{F}=1240$, and confirm that the theory scales well with cavity finesse.

The very nonlinear behavior of the phase-slope shows a clear optimum in absolute phase slope for input powers of about $8$~nW and a subsequent decrease in the absolute slope towards zero. While the phase slope is small for low powers due to the reduced saturation of the atoms, the saturation feature becomes power broadened for higher powers, once more leading to a reduction in the slope. The optimal phase slope is thus obtained for very low input powers, however, as we shall see below, this is not the optimal value for laser stabilization.

The full curve in FIG.~\ref{slope}~\textbf{a)} is a theoretical plot and we indicate a number of different input powers. At these powers we have performed scans of the atom-cavity spectrum and compared them to the theoretical model, in order to obtain a noise-free value for the phase slope at resonance. The fact that we see fluctuations of power-, atom number-, and technical noise or drift in the experiment is reflected by the misalignment between the dots and the theoretical behavior. 

Using the phase-slope it is possible to calculate the theoretically obtainable shot-noise limited (SNL) linewidth of a laser locked to the system. Here we find the minimal achievable linewidth by assuming that the detector efficiency is unity, and the lock is perfect. This can be found theoretically by using the expression \cite{Martin, ChristensenPHD}:
\begin{equation}
\Delta\nu=\frac{\pi h \nu}{2\eta_{\text{qe}}P_{\text{sig}}\left(\frac{d\phi}{d\nu}\right)^2}
\left(1+\frac{P_{\text{sig}}}{2P_{\text{ref}}}\right)
\end{equation}
Where $\frac{d\phi}{d\nu}$ is the phase slope at resonance, $P_{\text{sig}}$ is the carrier power and $P_{\text{ref}}$ is the reference power, which in our case is the power in the first order sidebands. $\eta_{\text{qe}}$ is the quantum efficiency of the detector which we assume here is one.

In FIG.~\ref{slope} \textbf{b)} we calculate this SNL linewidth $\Delta \nu$ and plot the curve corresponding to the slope of FIG.~\ref{slope} \textbf{a)}. We see that the optimum value of input power changes when we consider the SNL linewidth. For low powers the SNL linewidth increases dramatically as the shot noise starts to dominate the signal. This results in a relatively flat region around the optimum power spanning about an order of magnitude from $P_{\text{in}}\simeq10-100$~nW. The minimal value of $\Delta \nu$ is highly dependent on the ratio between sideband and carrier power. The optimal ratio of $\frac{P_{\text{ carrier}}}{2P_{\text{sideband}}}=1$ was used in these experiments. For these parameters we predict a minimal value of $\Delta \nu\approx40$~mHz which is comparable to the smallest laser linewidths ever reported \cite{Kessler,Bishof,Hafner}. By increasing the atom number it is possible to simultaneously decrease the projected linewidth of the locked laser, and increase the optimal operation power $P_{\text{in}}$.

\subsection{Dynamical range}
In FIG.~\ref{transition} the recorded signal $S_\Omega$ is shown for three different regimes where the maximal atomic phase shift is below, at, or above that corresponding to the maximal value of the transfer function. This shows the transition from a regime where the dispersion is largely unperturbed and represents the phase-response of the atoms well, to a regime where the response is significantly modified by the transfer function. 

At small phase-shifts we see a linear increase of the size of the signal proportional to the phase. At larger phase-shifts, however, the functional form of the cavity transfer function results in a mirroring effect of the dispersion signal for detunings above $\gamma_{\text{power}}$ where the phase shift is maximal. This has no influence on the slope around resonance, and will thus not affect the performance of an ideal frequency lock. It could, however, still limit the performance of a real servo system where the response time is not infinitely fast. 

We define the dynamical range of a lock to the dispersion signal as the range around resonance within which the dispersion \emph{slope} has constant sign. This range is dictated by the full width at half maximum (FWHM) of the power broadened transition linewidth. This corresponds to the width of the Lamb dip in the case of simple saturated absorption spectroscopy. The width, however, is modified by the slope of the Doppler broadened Gaussian dispersion feature. This dispersion causes line-pulling and thus decreases the dynamical range further. Lower temperatures will cause more pronounced line-pulling than higher temperature as the Doppler-broadened dispersion slope increases. While this effect actually causes a decrease in resonance slope it turns out that the fractional increase in the number of saturated (cold) atoms $N_{\text{sat}}$ outweighs this effect and the resonance slope is thus effectively increased for decreasing temperatures $T$. 

\begin{figure}[h]
	\includegraphics[width=\columnwidth]{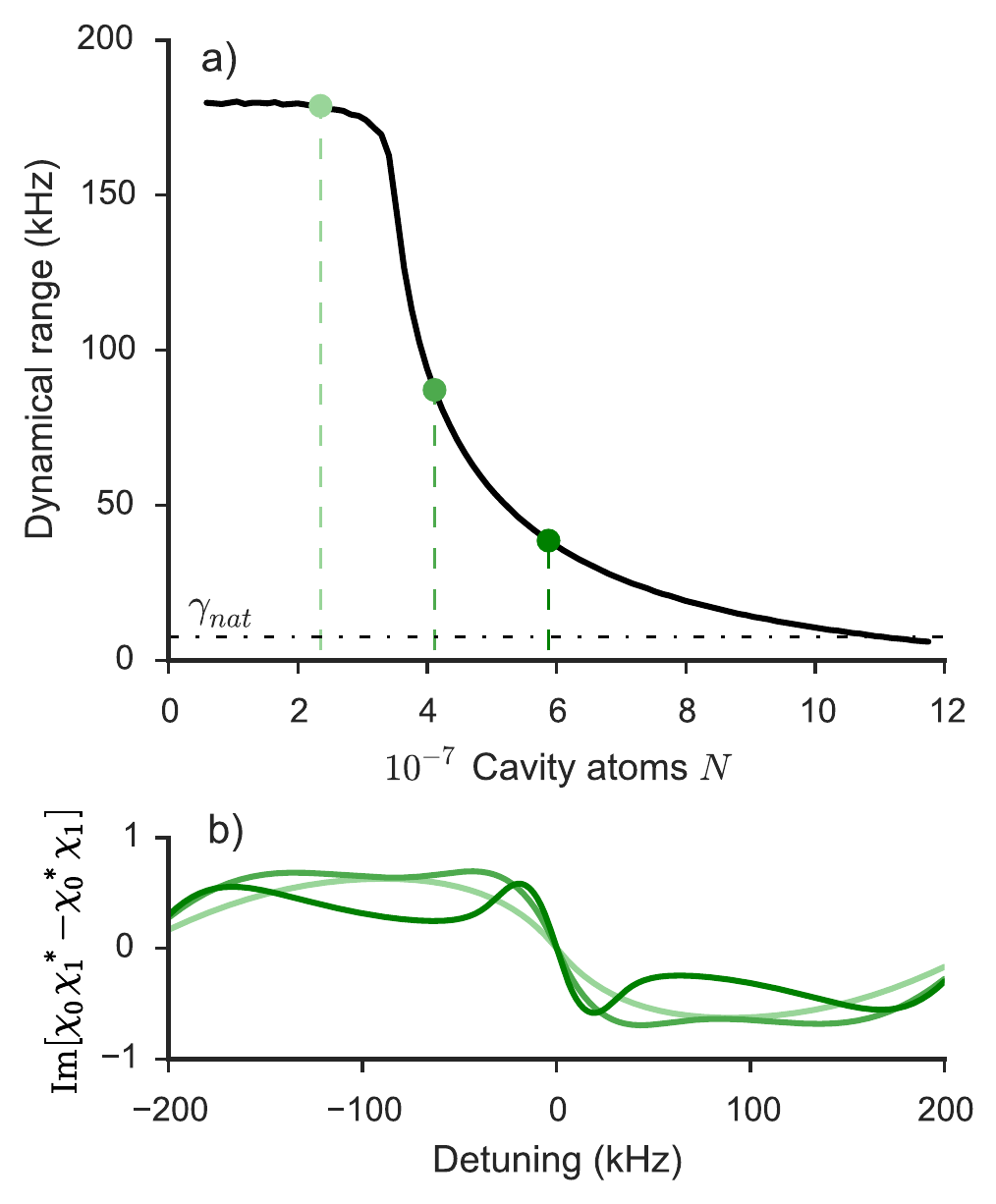}
	\caption{\textbf{a)} Dynamical range of the dispersion signal as a function of the intra-cavity atom number $N\times10^{-7}$. The dynamical range is the range around resonance where the sign of the dispersion slope is constant. The black dash-dotted line indicates the width of the dispersion if it was uniquely determined by the natural linewidth of the transition $\gamma_{\text{nat}}$. Three examples of the saturation dispersion are shown in \textbf{b)} and indicated by the dots on dashed lines in \textbf{a)}. The three examples correspond to atom numbers of $N=2.4\cdot10^7$ (light green), $N=4.1\cdot10^7$ (medium green), and $N=5.9\cdot10^7$ (dark green) respectively. The recorded scans in FIG.~\ref{transition} do not exceed $N=2\cdot10^7$ and are thus not limited by this effect. In \textbf{b)} the vertical axis, $\textrm{Im}[\chi_0\chi_1^*-\chi_0^*\chi_1]$, is the unit-less signal transfer function proportional to Eq.~\ref{eq:S0_1}, and is thus linear proportional to the recorded signal.  \label{dynamicalrange}}
\end{figure}

Finally the signal is modified by the cavity transfer function. Below the threshold in maximal phase-deviation set by this transfer function this is simply a phase-dependent scaling of constant sign and will thus not modify the dynamical range. Above this threshold, which becomes relevant in high $N$ systems such as the one reported here, we see a decrease of the dynamical range due to the slope-sign inversion dictated by the transfer function. A higher atom number $N$ increases the total phase, and thus pushes the system further beyond the threshold set by the transfer function boundaries. FIG.~\ref{dynamicalrange} shows the dependency on cavity atom number of the dynamical range for an in-coupling power of $P_{\text{in}}=100$~nW and a temperature of $T=2.5$~mK. This shows the initial dynamical range of $\Delta_{\text{dyn}}\simeq180$~kHz below threshold and a drop to few tens of kHz above the threshold. For typical atom numbers in our system we rarely exceed this threshold. For very high atom numbers, however, the range decreases asymptotically towards zero.

The dynamical range of a frequency locking scheme will be limited by the power-broadened transition linewidth $\gamma_{\text{power}}$ in all cases of FIG.~\ref{transition}. For higher atom numbers $N$, then, we will see another inversion within the narrow saturation dispersion, see FIG.~\ref{dynamicalrange}\,\textbf{b)}. Such an inversion will bring us into a regime where the dynamical range is limited by the properties of the transfer function $\chi$ rather than the power-broadened transition linewidth $\gamma_{\text{power}}$. Notice that this is only true if we require the sign on the slope to be constant. The sign of the signal itself will never change, and thus some degree of locking might still be possible for a flexible servo-system.

The dynamical range is of interest in particular regarding stability requirements for the interrogation laser. A standard requirement for the interrogation laser is that the interrogation laser linewidth should be smaller than the transition linewidth of the sample in order to resolve the line. If our initial interrogation laser linewidth is of the order of the natural linewidth ($\gamma=7.5$~kHz) this is well within the dynamical range below threshold. For very high atom numbers $N\gtrsim 1.1\times10^{8}$, however, the dynamical range decreases below the natural transition linewidth of the atoms. It is thus important that the interrogation laser is prestabilized to well within this dynamical range, before the atom-cavity error signal can be optimally utilized.

The aspects of the dynamical range considered here indicates that there is some optimal atom number depending on how efficient the servo can be made. While the slope around resonance increases linearly with the number of atoms $N$, and the dynamical range decreases severely above $N\approx2.5\cdot10^7$, an intermediate error signal could be preferable. Such a signal, like the intermediate (medium green) signal of FIG.~\ref{dynamicalrange}\,\textbf{b)}, provides the largest area under the error curve of the three shown. The preferred signal will depend on the particular experimental servo parameters.

\subsection{Locking condition effects}
Since our experimental realization is based on a cyclic operation, the cavity lock causes the length of the cavity to change dynamically throughout the experimental cycle. If the cavity dynamics is slower than required to obtain perfect locking, we see a small correction compared to the ideal locking signal of Eq.~\ref{eq:S0}. This causes large deviations in the DC transmission signal but has a relatively small effect on the phase response. When the cavity lock responsiveness is too slow the condition of constant resonance between the cavity and the laser carrier frequency will no longer be fulfilled. The atomic dispersion information will no longer be written onto the sideband frequencies but remains, in part or fully, on the carrier frequency. This means that $\chi_0$ is no longer purely real, and the dispersion term of the atomic phase shift affects the transmission. For high atomic phase-shifts, then, the transmission of the carrier component will be significantly reduced as the resonance condition is no longer necessarily fulfilled.

The locking condition determines some initial phase $\phi_{\text{init}}$ written onto the cavity phase \begin{equation}
\phi_{\text{cav}}=n\pi-\phi_{\text{init}}.
\end{equation}
Here we investigate three different cases. For the case of a fast cavity lock that can follow the system dynamics we have $\phi_{\text{init}}=\phi_{\text{D}}$ as shown in Eq.~\ref{eq:phicav}. A second idealized case is where the cavity lock is independent of the atoms inside the cavity $\phi_{\text{init}}=0$. This means that the length of the cavity simply follows the vacuum wavelength of the interrogation laser $L=n\frac{\lambda_{\text{vac}}}{2}$. The third, and the more realistic, case is where we have some perturbed phase due to the experimental conditions. In our case, the fact that the locking dynamics are relatively slow results in an initial phase given by the atoms under the influence of the cooling light $\phi_{\text{init}}=\phi_{\text{MOT}}$. The phase-shifts of the field components then becomes
\begin{eqnarray}
\phi_0 &=& n\pi +\phi_D + i\phi_A - \phi_{\text{init}}\\
\phi_{j} &=& (n+j)\pi - \phi_{\text{init}} \quad \text{for}\ j\neq0,
\end{eqnarray}
for some integer $n$.

First we look at the case of an atom-independent cavity lock. Here the cavity length ensures resonance with the laser beam assuming that there is only vacuum in the cavity. In this case the sidebands ($j\neq0$) are always resonant, but the carrier frequency ($j=0$) will be affected only by the atomic phase. In the limit of a very broad cavity linewidth $\kappa$ this situation is equivalent to having no active lock on the cavity length. The behavior under this condition thus gives us some insight into the case of a system operating in the deep bad cavity limit with stationary mirrors but resonant with the atomic transition. 

In the second case, relevant to our current system, a slow lock means that we lock to the atoms in the MOT while the cooling light is still on. The carrier frequency thus experiences some phase shift from the AC Stark shifted atoms ($\phi_{\text{MOT}}$), and this phase is written on the cavity length. Since the cavity cannot respond sufficiently fast to the subsequent conditions where MOT light is turned off, this modifies the phase of all $\chi_j$ with $\phi_{\text{MOT}}$. The phase-information from the non-perturbed atoms is now only on the carrier component. This heavily modifies the DC transmission, and also causes the antisymmetric behavior of the signal to be lifted as $\phi_{\text{MOT}}$ is not symmetric with respect to $\phi_{\text{D}}$. The carrier phase becomes
\begin{equation}
\phi_0 = n\pi - \phi_{\text{MOT}} +\phi_{\text{D}} + i\phi_{\text{A}}
\end{equation}
for integer $n$, and the sideband phases retain the phase written on the lock $\phi_{j\neq0}=(n+j)\pi - \phi_{\text{MOT}}$. In this case $\chi_0$ is no longer a purely real quantity, which modifies the signal. We have implemented this to first order by manually adding the measured phase-shift $\phi_{\text{MOT}}$ of the system to the transfer functions of the carrier and sideband frequencies. A full description must include the modified atom-light interaction in the cavity caused by this effective cavity detuning during the probing time.

\begin{figure}[h]
	\vspace{20pt}
	\includegraphics[width=\columnwidth]{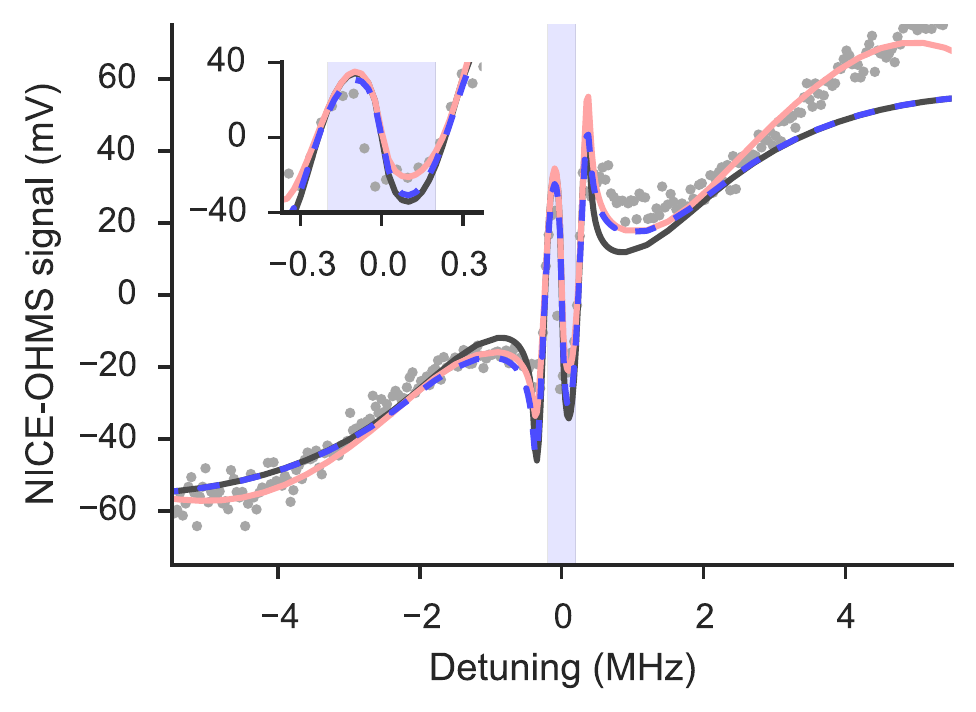}
	\caption{NICE-OHMS dispersion signal with asymmetry due to an AC Stark shift asymmetric with respect to the probe transition. Gray dots indicate data points, whereas the curves are theoretical plots using experimental system parameters. The black curve is plotted assuming a fast lock compared to the measurement dynamics, $\phi_{\text{init}}=\phi_{\text{D}}$, which is the optimal case of atom-cavity system locked to the carrier frequency of the laser. The dashed blue curve represents the case where the cavity is locked to resonance with the laser independently of the atoms, $\phi_{\text{init}}=0$. This primarily changes the absolute size of the phase. The light red curve includes a first-order correction for the AC Stark shifted atomic phase present in the cavity when the MOT beams are on $\phi_{\text{init}}=\phi_{\text{MOT}}$. This has large effects far away from resonance, but only little effect close to resonance. The parameters used are cavity atom number $N=2.5\cdot10^7$, temperature $T=2.8$~mK and an input power of $P_{\text{in}}=115$~nW. The light blue area marks the detuning range plotted in FIG.~\ref{dynamicalrange}\,\textbf{b)}.\label{cavity lock}}
\end{figure}

In FIG.~\ref{cavity lock} we show an example of a NICE-OHMS signal giving the dispersive response of the system. The NICE-OHMS signal has the expected features for a system with a large number of atoms in the cavity $N=2.5\cdot10^7$ where sharp features occur due to the limitations set by the transfer function. Three theoretical curves are plotted, which shows the theoretical behavior of the system assuming a fast cavity lock (black), a cavity locked independently of atoms (dashed blue), and a cavity locked to the AC Stark shifted atoms in the MOT (light red). While these different approaches only cause slight variations close to resonance, they are necessary to include in order to explain the signal for larger detunings. As expected, the features are slightly sharper in the case of a fast cavity lock.

\begin{figure}[h]
	\includegraphics[width=\columnwidth]{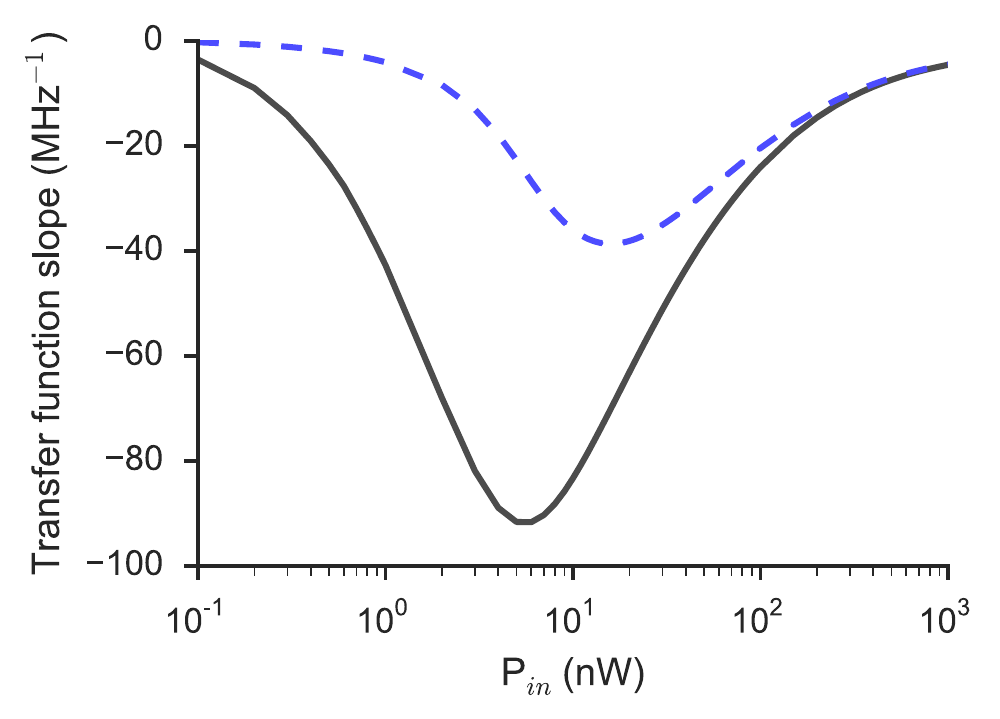}
	\caption{ Semi-logarithmic plot of the transfer function slope at resonance as a function of input power. The curves are theoretical slopes of the transfer function of the system. The dashed blue curve indicates the expected behavior if the cavity was locked to resonance with the laser independently of the atoms. The black curve is the expected behavior if the cavity lock is ideal, and the full cavity-atom resonance is tuned to the laser frequency. We see a decrease in slope for non-optimal cavity locking at all values of the input power $P_{\text{in}}$, as well as a distortion of the functional form which changes the position of the optimal slope. \label{locking}}
\end{figure}

The consequences of a non-optimal cavity locking condition on a laser lock is also considered here. FIG.~\ref{locking} shows two theoretical curves corresponding to optimal (fast) locking conditions (black), and atom-independent locking (dashed blue). In the case of optimal locking the system is close to a steady state. This can be realized either because the cavity lock is fast enough to follow the shift caused by turning off the trapping light, or by using a system operating in a continuous fashion. For the parameters used here ($T=2.5$~mK and $N=2.7\cdot10^7$) we see an optimal phase slope with the fast lock, for powers of about $P_{\text{in}}^{\text{opt}}=8$~nW. The phase slope is reduced for all values of the input power in the case of an atom-independent locking. The functional shape also changes, and the optimal input power is increased to about $P_{\text{in}}^{opt}=25$~nW. Notice that while the slope is definitely reduced, it is below a factor of two for powers relevant to laser locking. The optimal power also becomes more experimentally accessible, and the two cases are seen to give approximately identical slopes for powers larger than $P_{\text{in}}=400$~nW. This indicates that the performance of the cavity lock might not be of detrimental importance to the ultimate performance of the system within technically relevant parameter regimes.

\section{Conclusion}
We have experimentally investigated an atomic ensemble of cold $^{88}$Sr atoms in an optical cavity in the regime of high atomic phase-shift. The phase response of the system is recorded using the NICE-OHMS technique, and has promising features for frequency stabilization. 

The system operates in the bad cavity regime which suppresses the fluctuations caused by the finite temperature of the cavity. For the case of a narrow atomic transition, the bad cavity regime can still permit a high cavity finesse which yields a large number of photon round-trips. This causes the accumulated phase to grow beyond the approximately linear regime of the cavity transfer function, and mirroring effects of the phase-response can occur. These mirroring effects nonlinearly flips the slope of the dispersion signal around some maximal value. We experimentally mapped out the transition from the regime where the dispersion signal is an approximately linear representation of the atomic phase shift, to the regime where this representation is highly distorted by the cavity transfer function properties. We investigated the limitations this might have on an error signal for frequency locking of a laser. The mirroring effects causes a limitation of the dynamical range of a servo lock which must be included in the optimization of future servo systems operating using this technique.

We also investigated the ultimate performance of a laser stabilized to such a system and saw predictions consistent with earlier work \cite{Tieri}. These predictions rely on investigations of the phase slope achievable at resonance and do not take into account the limitations on a servo loop such as the dynamical range limitations that occur. We saw that the degradation of the signal slope caused by non-optimal cavity locking was not detrimental to the system and amounted to a factor of two for realistic experimental parameters. This means that even a slow cavity lock could produce promising results for laser stabilization, and opens for the possibility of leaving out the cavity lock entirely as long as the system is deep in the bad cavity regime.

\begin{acknowledgments}
We acknowledge support by ESA Contract No. 4000108303/13/NL/PA-NPI272-2012 and by the European Union through the project EMPIR 15SIB03 OC18.
\end{acknowledgments}

\appendix

\section{Theory}\label{Appendix:theory}
Here we give a very brief overview of the theory used to model the interaction of the light with the atom-cavity system. We follow \cite{Westergaard, Tieri} and model the system by using a Born-Markov master equation to describe the evolution of the system's density matrix $\hat{\rho}$. This evolution can be written as
\begin{equation}
\frac{d}{dt}\hat{\rho}=\frac{1}{i\hbar}\left[\hat{H},\hat{\rho}\right]+\hat{\mathcal{L}}[\hat{\rho}].
\end{equation}
The many-particle Hamiltonian describing the coherent evolution in a rotating interaction picture is given by 
\begin{eqnarray}
\hat{H}=&&\frac{\hbar\Delta}{2}\sum_{l=1}^{N}\hat{\sigma}_l^z+\hbar\eta\left(\hat{a}^\dagger+\hat{a}\right)\\\nonumber
&&+\hbar\sum_{l=1}^{N}g_l(t)\left(\hat{a}^\dagger\hat{\sigma}_l^-+\hat{\sigma}_l^+\hat{a}\right)
\end{eqnarray}
where $\Delta=\omega_a-\omega_c$ is the atom-cavity detuning, $\hat{\sigma}^{+,-,z}$ are the Pauli spin matrices and $\eta=\sqrt{\frac{2\pi\kappa P_{\text{in}}}{\hbar\omega_c}}$ is the classical drive amplitude. $\hat{a}$ and $\hat{a}^\dagger$ denote the annihilation and creation operators of the cavity mode respectively. The coupling rate between atoms and cavity is given by 
\begin{equation}
g_l(t)=g_0\cos(kz_l-\delta_lt)e^{-r_j^2/w_0^2},
\end{equation}
where $g_0$ is the vacuum Rabi frequency, $k$ is the wave number of the cavity mode, $z_l$ and $r_l$ denote the longitudinal and axial positions of the $l$'th atom, $\delta_l=kv_l$ is the Doppler shift contingent on the atom velocity $v_l$, and finally $w_0$ is the radial waist size of the cavity mode. Here the probing laser is assumed on resonance with the cavity at all times, $\omega_l=\omega_c$.

The incoherent evolution is described by the Liuvillian $\hat{\mathcal{L}}[\hat{\rho}]$ and is given by
\begin{eqnarray}
\hat{\mathcal{L}}[\hat{\rho}]=&-&\frac{\kappa}{2}\left\{\hat{a}^\dagger\hat{a}\hat{\rho}+\hat{\rho}\hat{a}^\dagger\hat{a}-2\hat{a}\hat{\rho}\hat{a}^\dagger\right\}\\\nonumber
&-&\frac{\gamma_{\text{nat}}}{2}\sum_{l=1}^{N}\left\{\hat{\sigma}_l^+\hat{\sigma}_l^-\hat{\rho}+\hat{\rho}\hat{\sigma}_l^+\hat{\sigma}_l^--2\hat{\sigma}_l^-\hat{\rho}\hat{\sigma}_l^+\right\}\\\nonumber
&+&\frac{1}{2T_2}\sum_{l=1}^{N}\left\{\hat{\sigma}_l^z\hat{\rho}\hat{\sigma}_l^z -\hat{\rho}\right\},
\end{eqnarray}
where $\kappa$ is the cavity decay rate, $\gamma_{\text{nat}}$ is the atomic transition linewidth and $\frac{1}{2T_2}$ is the inhomogeneous dephasing of the atomic dipole. The approach is thus based on a many-particle Hamiltonian $\hat{H}$ and a derived set of complex Langevin equations that includes the Doppler effect from the finite velocity of the atoms. The evolution is found by means of a Floquet analysis and solved for the steady state case. This will not be investigated further here, the interested reader is referred to \cite{Westergaard, Tieri}.

\end{document}